\documentclass[conference]{IEEEtran}

\IEEEoverridecommandlockouts    

\usepackage{cite}
\usepackage{amsmath,amssymb,amsfonts}
\usepackage{graphicx}
\usepackage{textcomp}
\usepackage{acro}
\usepackage{url}
\usepackage{multirow}
\usepackage{stfloats}
\usepackage[caption=false, font=footnotesize]{subfig}
\usepackage{algorithm}
\usepackage{algpseudocode}
\usepackage{hyperref}

\usepackage[table]{xcolor}

\usepackage{pgf} 
\usepackage{booktabs}

\newcommand{\heat}[1]{%
  \begingroup
  \pgfmathsetmacro{\v}{#1}%
  \pgfmathsetmacro{\vc}{min(15,max(-15,\v))}%
  \pgfmathtruncatemacro{\mix}{round((\vc+1)/30*100)}%
  \pgfmathtruncatemacro{\inv}{100-\mix}%
  \edef\colspec{white!\inv!green}%
  \expandafter\cellcolor\expandafter{\colspec}\strut #1%
  \endgroup
}

\title{\vspace{0.1in}\LARGE \bf Congestion Forecasting for Electric Vehicle Charging Scheduling with Fluid Queues}

\author{
	\parbox{\textwidth}{%
		\centering
		Joas Kahlert$^{1}$, Ruiting Wang$^{1}$, Jonas Mårtensson$^{1}$%
	}%
	\thanks{$^{1}$Decision and Control Systems, KTH Royal Institute of Technology, Stockholm, Sweden,
		{\tt\small kahlert@kth.se, ruiting@kth.se, jonas1@kth.se}}%
}

\DeclareAcronym{BET}{
  short = BET ,
  long  = Battery Electric Truck,
  short-plural = s 
}

\DeclareAcronym{E-VRP}{
  short = E-VRP ,
  long  = Electric Vehicle Routing Problem,
  short-plural = s 
}

\DeclareAcronym{HoS}{
  short = HoS ,
  long  = Hours of Service,
  short-plural = s 
}

\DeclareAcronym{SoC}{
  short = SoC ,
  long  = State of Charge,
  short-plural = s 
}

\DeclareAcronym{MIP}{
  short = MIP ,
  long  = Mixed Integer Program,
  short-plural = s 
}

\DeclareAcronym{E-VSP}{
  short = E-VSP ,
  long  = Electric Vehicle Scheduling Problem,
  short-plural = s 
}

\DeclareAcronym{PDF}{
  short = PDF ,
  long  = Probability Density Function,
  short-plural = s 
}

\DeclareAcronym{FCFS}{
  short = FCFS ,
  long  = First Come First Served,
  short-plural = s 
}

\DeclareAcronym{ODE}{
  short = ODE ,
  long  = Ordinary Differential Equation,
  short-plural = s 
}

\DeclareAcronym{EV}{
  short = EV ,
  long  = Electric Vehicle,
  short-plural = s 
}

\begin{document}

\maketitle

\thispagestyle{empty}
\pagestyle{empty}

\begin{abstract}
To support the adoption of electric transport systems, public charging opportunities are becoming increasingly important. In this dynamic environment, a central challenge for route planning and charging scheduling is forecasting charging-station availability under fluctuating demand. 

In this work, we propose a fluid-based forecasting method that accounts for uncertainty in both known and unforeseen electric vehicle arrival patterns while respecting station capacity constraints. We further evaluate the congestion forecasting method by applying it to an electric vehicle scheduling problem. 
Compared to scheduling frameworks that rely on standard baselines, charging schedules based on the fluid congestion forecasting model reduce waiting-related downtime by up to 14\%. Finally, we quantify how increased knowledge of vehicle arrivals and different levels of station congestion affect overall system performance.
\end{abstract}

\begin{keywords}
Electric vehicles, Queuing, Demand forecasting, Congestion, Charging scheduling, Mixed integer programming.
\end{keywords}

\section{Introduction}
\label{sec:introduction}

The electrification of the transport sector has grown significantly in recent years \cite{iea_2025_ev_charging}. To support large-scale adoption, increasing investments are being directed toward public charging infrastructure to provide reliable en-route charging opportunities, particularly for long-distance travel \cite{iea_2025_ev_charging}. With the infrastructure in place, the problem now shifts toward its efficient utilization.

In this context, many studies have addressed what can be summarized as the \ac{E-VSP}, which concerns the operational scheduling and charging management of \acp{EV} \cite{chau_2025}. Effective scheduling is essential to prevent both under- and over-utilization of charging infrastructure and to ensure reliable system operation \cite{wang_jochem_fichtner_2020, varshney_2025, sone_2024, alhanahi_2024}. 

For example, authors in \cite{varshney_2025} introduce a station control mechanism that regulates vehicle inflow via a feedback strategy informed by the current system state. \cite{wang_jochem_fichtner_2020} introduces a scenario-based uncertainty model for station demand profiles to determine optimal scheduling windows. Similarly, \cite{sone_2024} proposes an \ac{EV} charging scheduling algorithm that minimizes the number of simultaneous charging sessions. Their approach leverages a deep learning framework to predict vehicle arrivals, enabling charging operators to determine the appropriate number of charging sessions to order from the grid ahead of time.

In these works, an important aspect that has not been fully addressed is the forecasting of charging station availability, both at the time of planning and over a future time horizon. Route-planning applications for \acp{EV} in long-distance travel, such as ``A Better Route Planner'' \cite{abrp_2026}, and ``Chargemap'' \cite{chargemap_2026}, provide real-time information on charging-port availability, but rarely offer predictive forecasts. In line with this, \cite{ma_fang_2022} notes that most studies on \ac{EV} operational planning assume instantaneous and unconditional service. Accounting for future congestion is inherently difficult, as integrating realistic forecasting into operational models is often hindered by highly fluctuating and uncertain demand profiles.

Charging-planning models also typically assume either complete \cite{alhanahi_2024, wang2023b} or zero \cite{wang_jochem_fichtner_2020, sone_2024,varshney_2025} information exchange between stations and vehicles.
However, given the highly dynamic and complex nature of public charging systems involving multiple stakeholders, it is unrealistic to assume that all \acp{EV} are both willing and able to communicate their charging intentions within a global charging network. Conversely, without vehicle-to-station information exchange, availability predictions may become inaccurate or even infeasible.

In this study, we therefore construct an availability fluid-queuing-based forecast that explicitly accounts for uncertainty in \ac{EV} arrivals within a partial-communication framework. The proposed method accounts for both vehicles that communicate their arrival intentions and those that arrive without prior notice.

Fluid models have been widely adopted for approximations of discrete stochastic systems \cite{zychlinski_2023}. In contrast to discrete-state-space queuing models such as Markovian queues, the state and state transitions of fluid models are continuous and evolve according to a set of rate-based deterministic differential equations. Fluid models do not distinguish between discrete entities, which is particularly useful for information-sensitive environments that require customer anonymity. As will be shown later in this paper, the fluid model naturally decentralizes \acp{EV} charging interactions and complex queuing dynamics into aggregated station availability estimates on a rolling horizon. This architecture allows each \ac{EV} to compute its charging plans individually while still accounting for other \acp{EV} charging behavior.

In summary, we claim the following main contributions:
\begin{itemize}
    \item We developed a fluid queuing model to estimate charging station availability forecasts under congestion dynamics, filling the gap of integrating realistic, forward-looking congestion estimates into operational models that traditionally assume instantaneous and unconditional service. 
    \item We demonstrate the effectiveness of the proposed forecast model, which reduces waiting-related downtime by up to 14\% when compared to other benchmarks. 
    \item Applying the proposed model, we analyze station operating behavior and scheduling effectiveness in terms of planning improvements and schedule deviations, considering the effects of congestion level, station size, and knowledge of \ac{EV} arrivals.
\end{itemize}

\section{Methods}

\subsection{Problem Overview}
The objective is to develop a reliable availability forecast for a single, independently operating charging station that provides relevant information about the station's operating conditions to vehicles requesting data for route or charging planning.  For example, through applications such as \cite{abrp_2026, chargemap_2026} or other commercial planning software. In this context, station availability is characterized by two key indicators: the expected waiting time upon arrival before a charging session can begin, and the available charging power over the charging interval.

This information is particularly relevant for long-distance and en-route charging, where charging plans must be reliable, time-efficient, and dynamically updated. We therefore envision the availability forecast as an input to charging scheduling via a separate \ac{E-VSP}, whose formulation may vary with the specific application context. For this study, we use a general \ac{E-VSP} formulation to assess the predictive power of the proposed fluid forecasting model.

\subsection{Modeling \ac{EV} Arrivals}
\renewcommand{\arraystretch}{0.92} 

\begin{table}[htbp]
\centering
\caption{Notation Fluid Model}
\label{tab:notation}
\vspace{-8pt}
\begin{tabular}{p{0.3cm}|p{3.2cm}||p{0.3cm}|p{3.2cm}}
\hline
\textbf{} & Description & \textbf{} & Description \\
\hline
$q$ & \acp{EV} in queue
&  $\delta^*$ & Admitted load into service\\

$b$ & \acp{EV} in service
&$r$ & \ac{EV} departure rate\\

$E$ & Energy demand in queue  
& $K_n$ & \ac{EV} arrival PDF \\

$W$ & Energy demand in service 
& $\mu$ & \ac{EV} service rate \\

$w$ & Waited time into service 
& $\hat P^G$ & Maximum station power \\

$v$& Wait time into service
& $\hat P$ & Maximum charger power \\

$\lambda^u$ &  Scheduled \ac{EV} arrival rate 
& $P^S$ & Total supplied power \\

$\tilde\lambda$ &  Unscheduled \ac{EV} arrival rate 
& $P$& Supplied power per \ac{EV} \\

$\lambda$ & Aggregate \ac{EV} arrival rate 
& $\alpha$& Station congestion level \\

$\lambda^*$ & \ac{EV} admission rate into service 
& $\beta$ & Scheduled to unscheduled \ac{EV} ratio \\

$\delta^u$ & Arriving scheduled load  
& $c$ & Number of chargers \\

$\tilde\delta$ & Arriving unscheduled load 
&$\gamma$ & Service power cap\\

$\delta$ & Arriving aggregate load
&&\\

\hline
\end{tabular}
\end{table}

To fully leverage the available arrival information, we distinguish between two types of \acp{EV}:

\begin{enumerate}
    \item \textbf{Scheduled vehicles}: are willing to communicate and announce their charging intentions in advance. Although their exact arrival times and service requirements remain uncertain, their presence in the system is known beforehand.
    \item \textbf{Unscheduled vehicles}: are unwilling, unable, or unaware of communication and arrive without prior notice. Their arrivals are modeled statistically based on historical arrival rates.
\end{enumerate}




Assume that each scheduled \ac{EV} $n \in \mathcal{N}$ communicates its expected arrival time $\overline{t}_n$ and energy demand $E_n$ to the charging station. Since fluid models operate on rate functions rather than discrete processes, deterministic arrival instances must be translated into suitable rate-based representations. To this end, we model the arrival time of \ac{EV} $n$ as a normally distributed random variable, $t \sim \mathcal{N}\left(\overline{t}_{n}, \sigma_n^2\right)$, with \ac{PDF} $K_n(t):=K(t;\overline t_n, \sigma^2_n)$. Alternative arrival-time distributions can be incorporated without loss of generality.

To reflect increasing prediction uncertainty over the planning horizon, the standard deviation is defined as a function of the expected arrival time, $\sigma(\overline t_n)$. The form of $\sigma$ can either be statistically derived based on analyzing arrival patterns from historical data or communicated directly through the station-to-vehicle communication process for scheduled vehicles. For simplicity, we assume a linear relation of the form $\sigma(\overline t_n)=\varepsilon+k\overline t_n$.

The \ac{PDF} $K_n(t)$ can be interpreted as the instantaneous expected arrival-rate contribution of vehicle $n$. Integrating $K_n(t)$ over any time interval yields the probability of arrival within that interval. The expected aggregated arrival rate for the entire set of scheduled vehicles $\lambda^u(t)$ is given by the sum over $\mathcal N$. Similarly, the expected arriving energy demand for scheduled vehicles can also be represented by aggregating individual energy requirements $E_n$ around the time the workload is expected to arrive. 
\begin{equation}
    \label{eq:det_arrivals}
    \lambda^u(t)=\sum_{n\in\mathcal N}K_{n}(t),\quad \delta^u(t)=\sum_{n\in\mathcal N}E_{n}K_{n}(t).
\end{equation}

The arrival rates for the unscheduled \ac{EV} are defined based on historical rate functions $\tilde \lambda(t)$ and $\tilde\delta(t)$. Now the total arrival rates can conveniently be expressed by simply aggregating over both types of vehicles to get,
\begin{equation}
    \begin{aligned}
        \lambda(t)&=\tilde\lambda(t)+\lambda^u(t),\\
        \delta(t)&=\tilde\delta(t)+\delta^u(t).\\
    \end{aligned}
\end{equation}

More precisely, $\lambda(t)$ represents the expected arrival rate of all \acp{EV} at time $t$, measured in \acp{EV} per unit time. In contrast, $\delta(t)$ represents the expected rate at which required charging energy flows into the queuing system, measured in units of energy per unit time. Since this quantity is dimensionally equivalent to power, $\delta(t)$ can be interpreted as the power-equivalent charging load associated with the arriving \acp{EV}.

\subsection{The Fluid Queuing Model}
\begin{figure}[htbp]
\vspace{-6pt}
    \centering
    \includegraphics[width=1\linewidth]{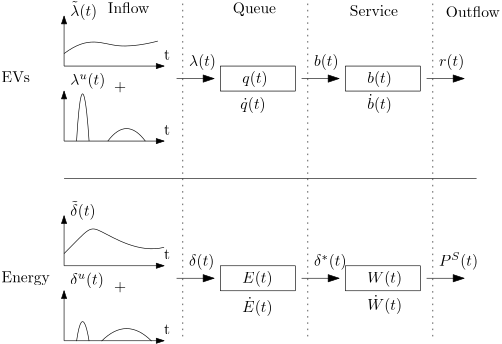}
    \caption{Coupled fluid queuing model.}
    \label{fig:queue_illustr}
\vspace{-6pt}
\end{figure}

We represent the charging station as a multi-server \ac{FCFS} queuing system, where arriving \acp{EV} are customers, chargers are servers, and charging sessions are service processes. For an overview of such systems, we refer to \cite{zychlinski_2023}; for a more in-depth analysis, see \cite{liu_whitt_2012}.

A schematic illustration of the queuing-to-service structure is provided in Figure~\ref{fig:queue_illustr}. In contrast to conventional fluid queuing models, our formulation tracks a coupled pair of fluid quantities that move through the system together: the vehicle flow and the corresponding energy-demand flow. Let $q(t)$ denote the number of \acp{EV} in queue, with total queued energy demand $E(t)$, and let $b(t)$ denote the number of \acp{EV} in service, with total remaining energy demand $W(t)$. The state-dependent rates of change of these quantities are denoted by the $\dot{\phantom a}$ operator.   

Both quantities undergo two distinct phases before leaving the system. Upon arrival, the fluid joins the tail of the queue in the waiting phase, where it must wait in typical \ac{FCFS} discipline before being admitted into the service phase. 

As incoming \acp{EV} have varying energy demand, the service time to complete their charging will also vary over time proportionally to the service rate $\mu(t)$. \acp{EV} leave service at the aggregate rate $r(t)$ equal to the rate of service per charger times the number of chargers that are currently occupied by the fluid in service $b(t)$. 
\begin{equation}
    r(t)=\mu(t)b(t).\label{eq:ev_out}\\
\end{equation}

Now, denote individual charger power by $\hat P$ and the current supplied power over all chargers by $P^S(t)$. Then, processing the aggregate demand $W(t)$ in service is done at rate,
\begin{equation}
    \mu(t)=\frac{P^S(t)}{W(t)},\label{eq:service_rate}
\end{equation}
per charger.

The total supplied power is equal to the charging power per charger times the number of active chargers and is bound by the maximum available power from the grid $\hat P^G$, 
\begin{equation}
    P^S(t)=\min(\hat P^G, \hat P b(t)).\label{eq:energy_out}\\
\end{equation}

If we define $\hat P^G=\gamma \hat Pc$, where $c$ is the number of chargers, then $\gamma$ is the ratio of the total power of all installed chargers to the maximum power supplied by the grid connection. In that case, $\gamma$ becomes an interesting parameter if the power connection to the grid is limited and cannot support many charging sessions at the same time.

Importantly, the queuing system operates in two distinct regimes, depending on whether the total number of \acp{EV} exceeds the service capacity $c$. The system is defined as \textbf{underloaded} when $b(t) < c$, and \textbf{overloaded} when $b(t) = c$. Note that $b(t) > c$ is physically impossible. The state-space equations for both regimes are given by a set of \acp{ODE}, see Figure \ref{fig:queue_illustr}.

In the \textbf{underloaded} state it holds that,
\begin{subequations}
    \label{eq:under}
    \begin{align}
        q(t)&=E(t)=w(t)=0,\label{eq:underq1}\\
        \lambda^*(t)&=\lambda(t),\quad \delta^*(t)=\delta(t).\label{eq:underq2}\\
        \dot b(t)&=\lambda^*(t)-r(t),\label{eq:unders1}\\
        \dot W(t)&=\delta^*(t)-P^S(t).\label{eq:unders2}
    \end{align} 
\end{subequations}

Equations ~\eqref{eq:underq1}-\eqref{eq:underq2} describe the flow dynamics within the queue. Because in the underloaded state, the queue is empty, arriving fluid enters into service immediately without waiting. Equation ~\eqref{eq:unders1}-\eqref{eq:unders2} describes the net rate of change of the fluid content in service.

In the \textbf{overloaded} state it holds that, 
\begin{subequations}
    \label{eq:over}
    \begin{align}
        b(t)&=c.\label{eq:overs1}\\
        \dot q(t)&=\lambda(t)-\lambda^*(t),\label{eq:overq1}\\
        \dot E(t)&=\delta(t)-\delta^*(t),\label{eq:overq2}\\
        \dot b(t)&=\lambda^*(t)-r(t)=0,\label{eq:overs2}\\
        \dot W(t)&=\delta^*(t)-P^S(t).\label{eq:overs3}
    \end{align}
\end{subequations}

Equation~\eqref{eq:overs1} states that the station is operating at full capacity $c$. Since the queue is nonempty in the overloaded regime, fluid enters and leaves the queue according to Eq.~\eqref{eq:overq1}-\eqref{eq:overq2}. While the number of \acp{EV} in service remains constant, implying a zero net rate of change Eq.~\eqref{eq:overs2}. However, this does not hold for the energy in service. Because \acp{EV} may have heterogeneous charging requirements, the total energy in service can increase or decrease depending on the requirements of \acp{EV} entering and leaving service, as in Eq.~\eqref{eq:overs3}. 

In the overloaded regime, fluid now has a nonzero queuing time. Let $v(t)$ be the time fluid has to \emph{wait} to enter service upon arriving in the queue. Also, let $w(t)$ be the time fluid has \emph{waited} in the queue upon entering service. Fluid that has arrived in service entered the queue $t - w(t)$ time units ago. Hence,
\begin{equation}
    w(t)=v(t-w(t)).\label{eq:wv}
\end{equation}

To this end, consider fluid that is just entering service at time $t$ with waited time $w(t)$. During this waiting period, additional fluid continues to accumulate in the queue according to the \ac{FCFS} discipline. By definition, the total amount of the accumulated fluid equals the current queue length. Over the time interval $[t - w(t), t]$, fluid arrives at rates $\lambda(t)$ and $\delta(t)$ respectively, which yields the relations,
\begin{equation}
    \label{eq:waiting_int}
    q(t)=\int^t_{t-w(t)}\lambda(s)ds,\quad E(t)=\int^t_{t-w(t)}\delta(s)ds.
\end{equation}

Assuming sufficient smoothness, we can solve Eq. \eqref{eq:under}, \eqref{eq:over}, and \eqref{eq:waiting_int} by reducing the size through substitutions, followed by any standard time stepping scheme with appropriately chosen initial conditions and state switching according to $q(t)\gtrless c$ in each timestep. In both regimes the unknowns are $q,b,E,W,\lambda^*,\delta^*,w$ matching the size of the system in Eq. \eqref{eq:under}, \eqref{eq:over}, and \eqref{eq:waiting_int}. The station availability indicators $P(t)$ and $v(t)$, can then be extracted from a simple postprocessing procedure by solving Eq. \eqref{eq:energy_out} and \eqref{eq:wv}.

\subsection{The \ac{E-VSP}}
To assess the applicability and accuracy of the fluid model, we employ a simplified scheduling problem in which station-availability forecasts inform charging decisions. We consider an \ac{EV} traveling along a fixed route with a sequence of charging opportunities. To reduce charging-related downtime, the \ac{EV} makes charging decisions using availability forecasts computed separately for each station.

We model the \ac{E-VSP} as a discrete-time \ac{MIP} optimization problem adopted from \cite{Kahlert2026}. For notation, see Table \ref{tab:evsp_notation}.
\begin{table}[htbp]
\centering
\caption{Notation \ac{E-VSP} Model}
\label{tab:evsp_notation}
\vspace{-6pt}
\begin{tabular}{p{0.32cm}|p{3.55cm}||p{0.32cm}|p{2.87cm}}
\hline
\textbf{} & Description & \textbf{} & Description \\
\hline
$x_i$ & Binary charging indicator
& $d_{i,j}$  & Road segment length \\

$e_i$ & Remaining energy upon arrival 
& $s_{i,j}$ & Road segment speed \\

$E_i$ & Energy charged at station
& $p$ & Energy consumption rate \\

$\overline t_i$ & Arrival time
& $C$ & \ac{EV} battery capacity\\

$t^*_i$ & Start of charging time 
&$\rho_i$& Charging overhead\\

$\underline t_i$ & Departure time 
&$t_k$& Discrete time\\

$\phi_{i,k}$ &Time-slot selector for arrival 
& $v_{i,k}$  & Waiting for charging\\

$\theta_{i,k}$ & Time-slot selector for charging
& $P_{i,k}$ & Available charging power\\

$\omega_i$ & Waiting time before charging 
& $\Delta t$& Temporal resolution\\

\hline
\end{tabular}
\end{table}

Consider an ordered set of charging stations $i\in\mathcal I$. Station $i^0$ and $i^e$ are the first and last stations on the corridor, while by the indexation $i^-/i^+$ we mean the previous or next station in order of station $i$. The set $\mathcal{I}^- = \mathcal{I} \setminus \{i^e\}$ contains all stations except the last one on the corridor. 

The waiting time and available charging power 
\begin{equation}
    v_{i,k}=v_i(t_k),\quad P_{i,k}=P_i(t_k),\label{eq:av_fun}
\end{equation}
are individually sampled for each station $i\in\mathcal I$ using the forecast estimates from the fluid model at discrete times $t_k$.

The charging and discharging constraints are  
\begin{subequations}
    \label{eq:charge}
    \begin{align}
        & e_{i^+}=e_i-d_{i,i^+}p+ E_i, &\forall i\in \mathcal I^-,\label{eq:char_event1}\\
        &E_i\leq x_iC,&\forall i\in\mathcal I,\label{eq:char_event2}\\
        & \eta C\leq e_i\leq C, &\forall i\in \mathcal I,\label{eq:SoC}\\  
        & E_i \leq \Delta t\sum_{k\in\mathcal K}\theta_{i,k}P_{i,k},&\forall i\in \mathcal I,\label{eq:energy_bound_1}\\
        &  0\leq E_i\leq C-e_i,&\forall i\in\mathcal I,\label{eq:energy_bound_3}
    \end{align}
\end{subequations}
where Eq. \eqref{eq:energy_bound_1} integrates the available charging power over the discrete charging interval, equaling the charged energy over that interval.

The time-space constraints are 
\begin{subequations}
    \label{eq:times}
    \begin{align}
        & t^*_i\geq \overline t_i+\omega_i+\rho_ix_i,&\forall i\in \mathcal I,\label{eq:visit1}\\
        & \underline t_i\geq t^*_i+\Delta t\sum_{k\in\mathcal K}\theta_{i,k},&\forall i\in \mathcal I,\label{eq:visit2}\\
        &\overline t_{i^+}\geq \underline  t_i+\frac{d_{i,i^+}}{s_{i,i^+}},&\forall i\in \mathcal I^-,\label{eq:next_station}   
    \end{align}
\end{subequations}
where Eq. \eqref{eq:visit2} expresses the aggregated length of the charging interval. 

The indicator constraints for arrival and charging are
\begin{subequations}
    \label{eq:wating}
    \begin{align}
        &\theta_{i,k}=1,\quad\text{if}\quad t_k\in[t^*_i,\underline t_i],&\forall (i,k)\in\mathcal I\times\mathcal K,\label{eq:theta_ind}\\
        &\phi_{i,k}=1,\quad\text{if}\quad \overline t_i\in[t_k,t_{k+1}],&\forall (i,k)\in\mathcal I\times\mathcal K^-,\label{eq:phi_ind}\\
        &\omega_i = \sum_{k\in\mathcal K}\phi_{i,k}v_{i,k},&\forall i\in\mathcal I,\label{eq:omega_ind}\\
        &\sum_{k\in\mathcal K}\phi_{i,k} = x_i,&\forall i\in\mathcal I,\label{eq:xphi}\\
        &\phi_{i,k},\ \theta_{i,k}\leq x_i,&\forall (i,k)\in\mathcal I\times\mathcal K,\label{eq:tighten}
    \end{align}
\end{subequations}
where, Eq. \eqref{eq:theta_ind} activates the charging indicator only between the start of charge and the time of leaving the station; Eq. \eqref{eq:phi_ind} activates the arrival indicator in the correct time-slot; Eq. \eqref{eq:omega_ind} selects the correct waiting time for the arrived time slot; Eq. \eqref{eq:xphi} ensure at least one arriving time is selected if the station is visited; and Eq. \eqref{eq:tighten} tightens the overall binary formulation. 

We note that the formulation in Eq.~\eqref{eq:wating} can be implemented efficiently in modern \ac{MIP} solvers using indicator constraints, or alternatively, can be linearized conventionally through appropriately defined auxiliary variables and big-M formulations.

For completeness, the variable domains are 
\begin{equation}
    \label{eq:domains}
    \begin{aligned}
        &e_i,\ E_i,\ \overline t_i,\ t^*_i,\ \underline t_i,\ \omega_i\in\mathbb R^+,&\forall i\in\mathcal I,\\
        &x_i,\ \phi_{i,k},\ \theta_{i,k}\in\{0,1\},&\forall (i,k)\in\mathcal I\times\mathcal K. 
    \end{aligned}
\end{equation}

The most efficient charging schedule is the one that will result in the fastest travel time through the corridor $\overline t_{i^e}$, which is equivalent to minimizing the total travel downtime related to waiting and charging,
\begin{equation}
    \tau=t_{i^e}-\sum_{i\in I^-}\frac{d_{i,i^+}}{v_{i,i^+}}.
\end{equation}

The full \ac{MIP},
\begin{equation*}
\begin{aligned}
    \min\quad &\tau,\\
    \text{s.t.}\quad &\eqref{eq:charge},\ \eqref{eq:times},\ \eqref{eq:wating},\ \eqref{eq:domains},\\
\end{aligned}
\end{equation*}

is solved with Gurobi LLC \cite{gurobi}. 

\subsection{Evaluating the Performance}\label{sect:eeee}

We focus on three most important system parameters for performance evaluation: $c$, the charging station capacity (i.e., the number of chargers); $\alpha$, the average congestion level; and $\beta$, the fraction of scheduled vehicles.

To represent some fluctuating arrival rate over the time of day, we assume a simple periodic unit baseline arrival rate $f^{\text{sin}}(t)$. The total arrival rate is then scaled according to the maximum throughput of vehicles $R$, resulting in $f(t) = R \alpha f^{\text{sin}}(t)$. With this arrival rate we expect the station to operate at $100\alpha\%$ capacity on average. The unscheduled arrival rate is given by $\tilde\lambda(t) = (1-\beta) f(t)$, while scheduled arrivals $\overline{t}_n$ are sampled from the remaining rate $\beta f(t)$ and aggregated according to Eq.~\eqref{eq:det_arrivals} to obtain $\lambda^u(t)$. The arriving workload is derived analogously. This means that the aggregate arrival rates $\lambda(t)$ and $\delta(t)$ still follow the baseline arrival-rate profile, but they are adjusted using more precise information about when \acp{EV} arrive, increasing with $\beta$.

The process for evaluating the fluid model with respect to specific system parameters is summarized in Algorithm \ref{alg:algorithm}. For further details, we refer to our \href{https://github.com/joask9/fluidq.git}{GitHub repository.}

The performance evaluation begins by initializing the key parameters $\alpha, \beta, c$. Each specific parameter set is then evaluated over multiple scenarios $\ell\in\mathcal L^{scn}$. Each scenario is defined by a slightly randomized environment for a more general assessment of different problem settings. This includes randomization for station-specific overhead $\rho$, the baseline arrival rate $f^{\text{sin}}(t)$, and the initial state of charge of the optimized vehicle.

For each scenario, we generate three different station availability estimates $(v,P)$: 1) using the fluid model; 2) using a Monte Carlo simulation in which scheduled arrivals are treated as exact arrivals without deviation; and 3) without estimates, assuming that the station is never congested. We refer to these cases as \emph{fluid}, \emph{simulation}, and \emph{naive}, respectively. Based on each availability estimate using Eq.~\eqref{eq:av_fun}, the vehicle determines its optimal charging strategy, namely the charging location $x$ and charging amount $E$, by solving the \ac{E-VSP}.

Each case yields the minimum predicted downtime under its respective station availability assumption. However, the resulting optimal charging strategies may differ substantially depending on the underlying performance estimate.

To assess their practical effectiveness, we evaluate each charging strategy in a separate Monte Carlo simulation. In this evaluation setting, the vehicle follows its predetermined charging plan, while station-side arrivals are stochastic according to their respective rate functions, rather than treated as exact arrivals as in case 2). This simulation is intended to reflect real-world environment where actual station behavior may evolve differently from the forecasts available at the time of scheduling, in which each strategy is tested independently.

For each evaluation, we record the experienced downtime. Ideally, the realized downtime should be close to the optimal downtime predicted by the \ac{E-VSP}. However, due to the differing assumptions underlying the station availability estimates, the strategies may exhibit varying levels of performance.

\begin{algorithm}
\caption{Model evaluation process, (Fluid (F), Simulation (S), Naive (N)).}
\label{alg:algorithm}
\begin{algorithmic}[1]

\State $\alpha, \beta, c\gets $ initialize key system parameters
\ForAll{$\ell\in\mathcal L^{scn}$}
    \ForAll{$i \in \mathcal I$}
        \State $\lambda_{\ell,i}, \delta_{\ell,i} \gets$ \Call{GenInput}{$\alpha,\beta,c$}
        \State $v_{\ell,i}^{F} ,P_{\ell,i}^{F}  \gets$ \Call{SolveFluid}{$\lambda_{\ell,i}, \delta_{\ell,i}$}
        \State $v_{\ell,i}^{S} , P_{\ell,i}^{S}  \gets$ \Call{SimDiscrete$_\ell$}{$\lambda_{\ell,i},\delta_{\ell,i}$}
        \State $v_{\ell,i}^{N} , P_{\ell,i}^{N}  \gets 0, \hat P$
    \EndFor
    \State $x_\ell^{F}, E_\ell^{F}\gets$ \Call{SolveMIP}{$ v_{\ell,i}^{F}, P_{\ell,i}^{F}$}
    \State $x_\ell^{S},E_\ell^{S}\gets$ \Call{SolveMIP}{$v_{\ell,i}^{S}, P_{\ell,i}^{S}$}
    \State $ x_\ell^{N}, E_\ell^{N}\gets$ \Call{SolveMIP}{$ v_{\ell,i}^{N}, P_{\ell,i}^{N}$}
    \State $\tau_{\ell}^{F}\gets$ \Call{EvlSolution$_\ell$}{$ x_\ell^{F}, E_\ell^{F}$}
    \State $\tau_{\ell}^{S}\gets$ \Call{EvlSolution$_\ell$}{$x_\ell^{S},E_\ell^{S}$}
    \State $\tau_{\ell}^{N}\gets$ \Call{EvlSolution$_\ell$}{$x_\ell^{N}, E_\ell^{N}$}
\EndFor
\State \Return  $\tau^{F}, \tau^{S},\tau^{N}  \gets \text{mean}(\tau_{\ell}^{F}) ,\ \text{mean}(\tau_{\ell}^{S}) ,\ \text{mean}(\tau_{\ell}^{N})$
\end{algorithmic}
\end{algorithm}

\section{Results and Discussion}

\subsection{Fluid Queuing Model Behavior}
To roughly indicate the computational performance of the proposed method, we report the observed runtimes without providing a detailed benchmark of parameter settings, problem sizes, or hardware specifications. The individual instances are solved quickly: the fluid-model time-stepping scheme runs in approximately $0.01$–$0.05$ seconds, while Gurobi solves the \ac{E-VSP} in about $0.5$–$10$ seconds.

\begin{figure}[thbp]
    \centering
    \vspace{-5pt}
    \hspace*{-0.4cm}
    \includegraphics[width=1.05\linewidth]{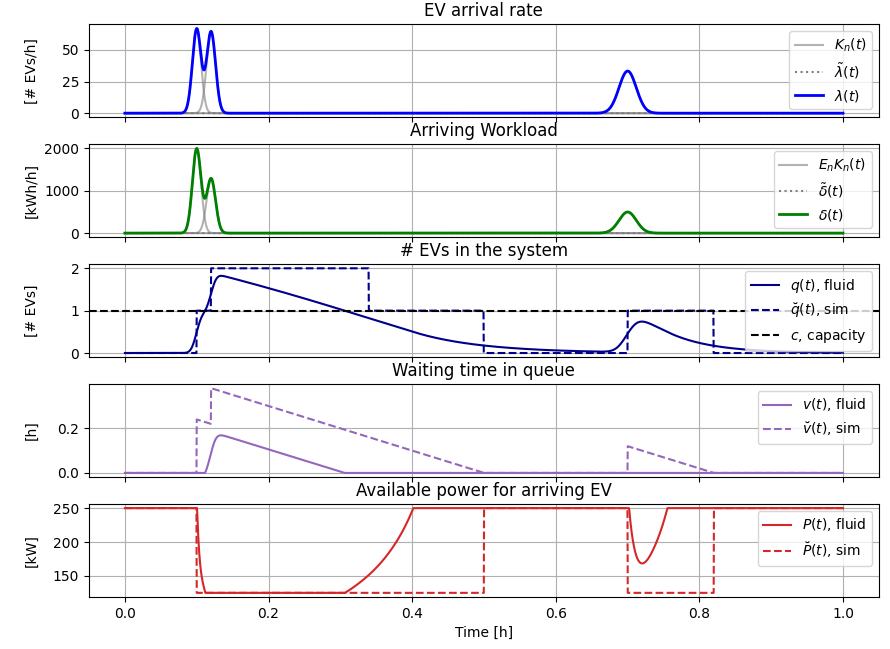}
    \caption{Illustration of fluid queuing model and discrete simulation comparison at a single station with one charger. Three vehicles are expected to arrive at times $0.1, 0.12, 0.7$, with energy requirements $30, 20, 15$ kWh. In the simulation, \acp{EV} are treated as separate discrete entities, whereas in the fluid model, vehicles are treated as one fluid quantity.}
    \label{fig:example}
    \vspace{-10pt}
\end{figure}

We first illustrate the behavior of the fluid queuing model in comparison to the simulation using a toy example shown in Fig.~\ref{fig:example}, featuring three scheduled arrivals and no unscheduled arrivals.

Subplots~1-2 exhibit three distinct peaks centered at the expected arrival times, showing the vehicle arrival rate and the corresponding energy demand. Since the third vehicle is scheduled further in the future, its arrival time is subject to greater uncertainty, according to its \ac{PDF}.

Based on these arrival rates, the fluid queuing model and the simulation yield the availability forecast (subplots~3–5) in terms of predicted number of \acp{EV}, waiting time upon arrival, and available charging power. While the simulation treats arrivals as discrete deterministic events, the fluid model operates across continuous values while capturing the underlying uncertainties in arrival patterns.

As a result, when the system is near the congestion threshold, and it is uncertain whether an arriving \ac{EV} will need to wait, the model does not impose a binary waiting event. Instead, as shown in Subplots~4-5, the available charging power is reduced below its nominal level, representing congestion through a continuous reduction in service rate. 

The behavior for instances of larger scale is more intricate, but theoretically no different, see the following figure.  

\begin{figure}[h!]
    \centering
    \vspace{-6pt}
    \hspace*{-0.4cm}
    \includegraphics[width=1.05\linewidth]{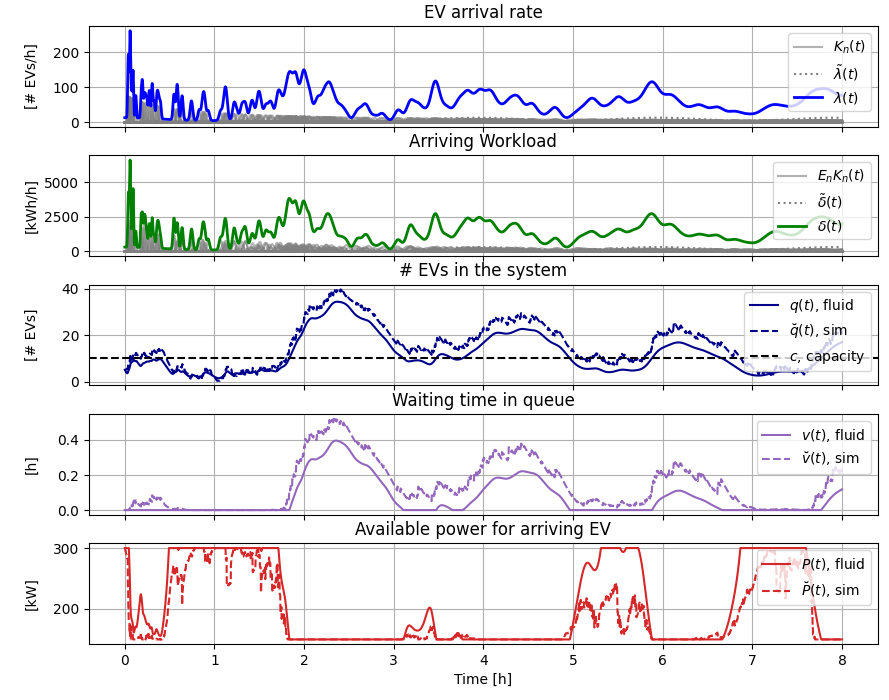}
    \caption{Illustration of a larger fluid queuing model with $10$ chargers and 50-50 scheduled-unscheduled arrivals. Fluid station estimators are closer to the simulated estimations compared to the smaller system in Fig. \ref{fig:example}.}
    \label{fig:example_big}
    \vspace{-7pt}
\end{figure}

\subsection{Evaluation of the E-VSP}

Through the evaluation process described in Section \ref{sect:eeee} we quantify the potential gain in downtime reduction achieved by the charging plan derived from the fluid model, compared to both the discrete simulation-based plan and a naive benchmark. The mean improvement is reported in Table \ref{tab:downtime} and highlighted in green to indicate superior performance. The fluid model gives almost exclusively better results than the other two approaches. 

An \ac{EV} following the naive estimation strategy only minimizes the known station-specific overhead $\rho_i$, which includes factors such as detour time. Intuitively, this is a reasonable approach, as users tend to choose the most convenient charging options when planning. However, this strategy is only effective when the expected congestion level is not too high. The results show that, for station congestion levels above approximately $50\%$, planning without availability forecasts can lead to significant downtime increases of up to $14\%$. This result suggests that public charging stations should either be designed to operate at an average congestion level below $50\%$, or be supported by a communication framework to avoid significant scheduling overhead for \ac{EV} charging.

The performance gap between the fluid-based and naive strategies can be interpreted as the value of information sharing, that is, the benefit of communicating charging intentions. This provides an interesting perspective on the conditions under which communication is most valuable. Relative to the naive approach, communication frameworks are especially important if the station is expected to be congested often. In addition, smaller stations tend to benefit more from providing an availability forecast.

The fluid model also predominantly outperforms the discrete state-space simulation. By accounting for stochastic uncertainty and time-horizon estimation confidence, the properties of the fluid model’s enable \acp{EV} to make uncertainty-aware decisions that are overlooked in the deterministic simulation approach, thereby improving the overall robustness of charging schedules.

\setlength{\tabcolsep}{3pt}
\begin{table*}[t]
\centering
\caption{Improvement (\%) of relative downtime to Naive and Simulation baselines.}
\label{tab:downtime}
\small

\resizebox{\textwidth}{!}{%
\begin{tabular}{c|
cccccc|
cccccc||
cccccc|
cccccc}
& \multicolumn{6}{c|}{$\alpha$ (Naive)}
& \multicolumn{6}{c||}{$\beta$ (Naive)}
& \multicolumn{6}{c|}{$\alpha$ (Sim)}
& \multicolumn{6}{c}{$\beta$ (Sim)} \\
\hline
$c$
& 0.1 & 0.5 & 0.7 & 0.9 & 1.0 & 1.1
& 0.01 & 0.2 & 0.4 & 0.6 & 0.8 & 1.0
& 0.1 & 0.5 & 0.7 & 0.9 & 1.0 & 1.1
& 0.01 & 0.2 & 0.4 & 0.6 & 0.8 & 1.0 \\
\hline
2
& \heat{-0.07} & \heat{2.09} & \heat{8.55} & \heat{10.71} & \heat{14.14} & \heat{10.83}
& \heat{5.27} & \heat{2.28} & \heat{7.05} & \heat{9.49} & \heat{14.55} & \heat{10.04}
& \heat{0.27} & \heat{3.93} & \heat{0.58} & \heat{-0.89} & \heat{2.95} & \heat{1.21}
& \heat{3.15} & \heat{4.16} & \heat{3.12} & \heat{1.68} & \heat{-0.73} & \heat{6.28} \\
5
& \heat{0.00} & \heat{0.74} & \heat{4.98} & \heat{10.67} & \heat{10.89} & \heat{12.61}
& \heat{4.76} & \heat{6.36} & \heat{5.24} & \heat{8.80} & \heat{13.22} & \heat{11.70}
& \heat{0.06} & \heat{1.79} & \heat{3.93} & \heat{1.86} & \heat{3.22} & \heat{1.23}
& \heat{2.67} & \heat{0.76} & \heat{2.44} & \heat{2.98} & \heat{1.86} & \heat{2.91} \\
10
& \heat{-0.00} & \heat{0.13} & \heat{3.23} & \heat{9.01} & \heat{13.35} & \heat{9.77}
& \heat{5.31} & \heat{6.44} & \heat{9.04} & \heat{11.67} & \heat{12.09} & \heat{12.17}
& \heat{-0.00} & \heat{0.90} & \heat{1.97} & \heat{1.37} & \heat{3.76} & \heat{1.22}
& \heat{3.10} & \heat{2.73} & \heat{1.32} & \heat{0.36} & \heat{-0.75} & \heat{0.40} \\
50
& \heat{0.00} & \heat{0.01} & \heat{1.77} & \heat{10.67} & \heat{12.18} & \heat{9.14}
& \heat{9.09} & \heat{8.37} & \heat{11.17} & \heat{7.39} & \heat{12.06} & \heat{11.14}
& \heat{0.00} & \heat{0.32} & \heat{1.54} & \heat{0.31} & \heat{3.31} & \heat{2.77}
& \heat{4.77} & \heat{2.58} & \heat{2.10} & \heat{1.05} & \heat{2.71} & \heat{1.53} \\
\hline
\end{tabular}%
}
\end{table*}

\subsection{Key Performance Parameters}
\begin{figure*}[htbp]
\vspace{-6pt}
    \centering
    \subfloat[Downtime vs. Scheduled Vehicle Ratio $\beta$ ($\alpha = 0.9$)]
    {\includegraphics[width=\textwidth]{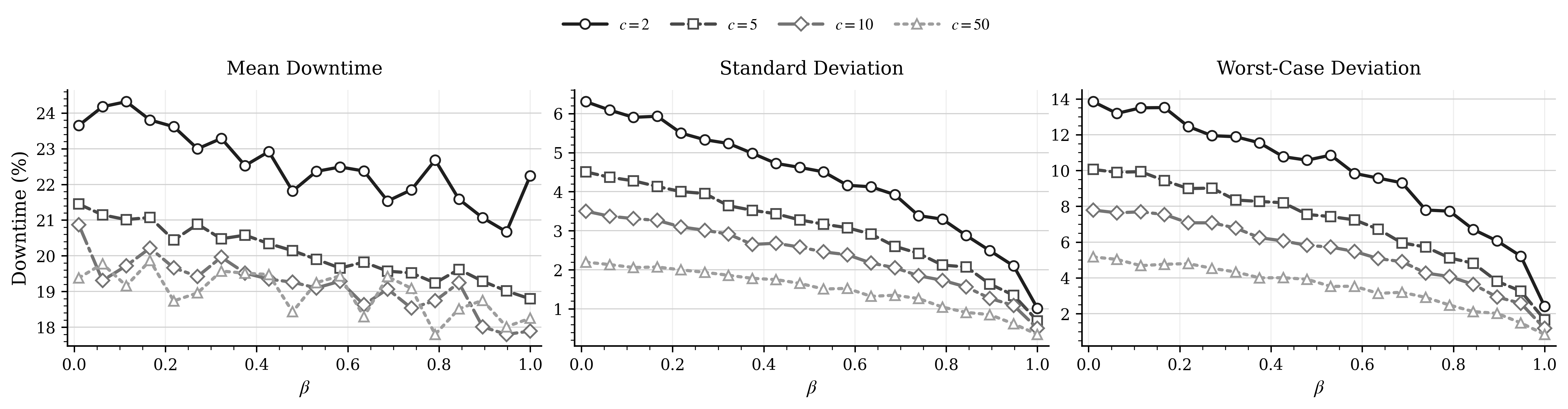}}\\
    \subfloat[Downtime vs. Congestion Intensity $\alpha$ ($\beta = 0.7$)]
    {\includegraphics[width=\textwidth]{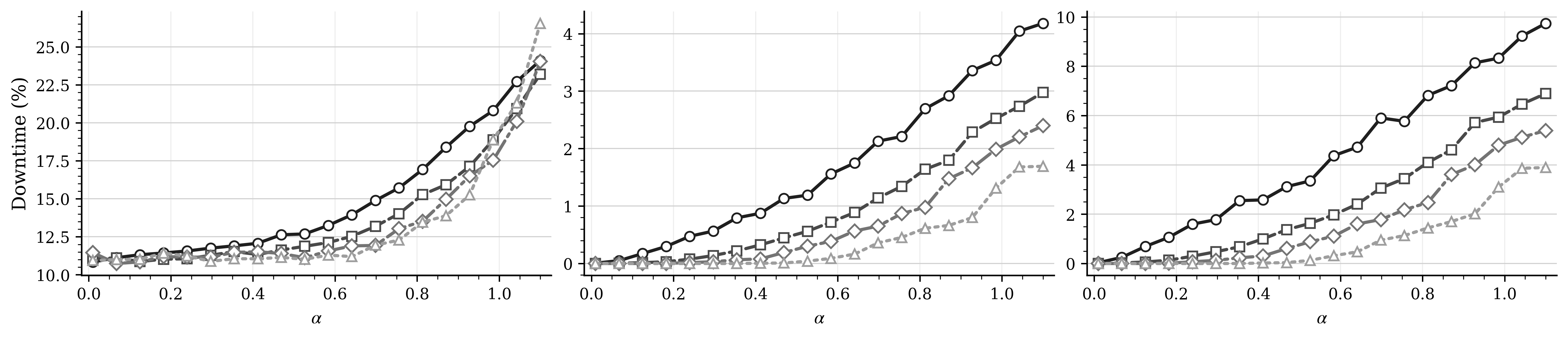}}
    \caption{\ac{E-VSP} downtime over key system parameters. The top row illustrates the impact of the scheduled vehicle ratio ($\beta$) at a fixed congestion level, while the bottom row depicts the impact of congestion intensity ($\alpha$) at a fixed scheduling ratio.}
    \label{fig:alphabeta}
\vspace{-6pt}
\end{figure*}

Moving on to further analyzing the effects of charging scheduling over the different key system parameters, we now solely focus on charging planning through the fluid model. As before, we evaluate system efficiency over different scenarios characterized by the congestion level $\alpha$, the fraction of scheduled vehicles $\beta$, and the station capacity/size $c$. The results are visualized in Fig. \ref{fig:alphabeta}.

To exemplify how to interpret the graphs in Figure~\ref{fig:alphabeta}, consider a scenario where $10\%$ of \acp{EV} communicate with stations part of the \ac{E-VSP} ($\beta=0.1$). For a station with five chargers, this implies that the scheduled \ac{EV} experiences, on average, 21\% downtime, with a standard deviation of 4.3\%. In worst-case scenarios, downtime may increase by an additional 10 percentage points, resulting in total downtime of up to approximately 31\%.

Figure~\ref{fig:alphabeta} (a) shows that downtime follows an approximately linear trend with respect to the ratio of information-sharing vehicles, $\beta$, decreasing by about $2$--$3\%$ across all station sizes. This corresponds to a relative reduction of roughly $10$--$13\%$. More notably, both the standard deviation and the worst-case deviation per scenario decrease substantially as the proportion of information-sharing \acp{EV} increases. In addition, downtime generally decreases as station size increases, further indicating that information sharing is particularly beneficial for smaller stations with fewer chargers.

Figure~\ref{fig:alphabeta} (b) shows a pronounced increase in downtime as the congestion level $\alpha$ rises, ranging between $11$--$25\%$. Congestion also negatively affects variability, increasing both the standard deviation and the worst-case deviation.

\section{Conclusion}

A primary source of uncertainty for public charging availability is the fluctuating demand and uncertainty in \ac{EV} arrivals. This study employs a fluid queuing model to forecast charging station availability in terms of expected waiting time and available charging power.

For effective scheduling of \acp{EV}, results show that vehicle-to-station communication becomes especially relevant for smaller charging stations with fewer chargers. In such cases, having access to station availability information can reduce waiting-related downtime by up to 14\%. Our results also suggest that stations operating above an average congestion level of $50\%$ benefit significantly from vehicle-to-station communication, which reduces the risk of substantial \ac{EV} charging overhead. Furthermore, increasing access to \ac{EV} arrival information enables more efficient and robust charging schedules in terms of standard deviation and worst-case outcomes, while also reducing the average experienced downtime by 10-13\%.

\bibliographystyle{IEEEtran}
\bibliography{root} 
	
\end{document}